# Dispersion coefficients for the interaction of inert gas atoms with alkali and alkaline earth ions and alkali atoms with their singly ionized ions

Sukhjit Singh[a], Kiranpreet Kaur[a], B. K. Sahoo[b*] and Bindiya Arora[a]
[a]Department of Physics, Guru Nanak Dev University, Amritsar, Punjab-143005, India
[b]Theoretical Physics Division, Physical Research Laboratory, Navrangpura, Ahemadabad-380009, India
e-mail[*] : bijaya@prl.res.in

___________________________________________________________________________________________

We report the dispersion coefficients for the interacting inert gas atoms with the alkali ions, alkaline earth ions and alkali atoms with their singly charged ions. We use our relativistic coupled-cluster method to determine dynamic dipole and quadrupole polarizabilities of the alkali atoms and singly ionized alkaline earth atoms, whereas a relativistic random phase approximation approach has been adopted to evaluate these quantities for the closed-shell configured inert gas atoms and the singly and doubly ionized alkali and alkaline earth atoms, respectively. Accuracies of these results are adjudged from the comparison of their static polarizability values with their respective experimental results. These polarizabilities are further compared with the other theoretical results. Reason for the improvement in the accuracies of our estimated dispersion coefficients than the data listed in [At. Data and Nucl. Data Tables **101**, 58 (2015)] are discussed. Results for some of the atom-ion interacting systems were not available earlier, these results and the other reported improved results will be very useful for the comprehensive understanding of the collisional physics involving these atom-atom and atom-ion interactions in the cold atom and atom-ion hybrid trapping experiments at the low-temperature regime.

___________________________________________________________________________________________

**1 Introduction**

The long-range interactions between the atoms and molecules play prominent roles in the low-energy and low-temperature collision experiments [1]. Thus, these interactions are expedient for understanding atomic collision physics that are essential for guiding the laser cooling and trapping techniques of atomic systems, in the photoassociation spectroscopy and for analysing the magnetic field induced Feshbach resonances [2-6]. In addition, these interactions are instrumental in the chemical processes for the charge-exchange and molecule formation at the single particle level. Comprehensive understanding about behaviour of these interactions is very useful in explaining various quantum phase transitions [7], quantum computing techniques [8], endowing continual atom-ion sympathetic cooling mechanisms [9, 10] etc. In fact, investigations of atom-ion interactions have drawn recent attention of the researchers for several reasons to carry out many inventive studies [9, 11, 12]. For example, several applications of co-trapped atoms and ions at the low energy scale have been demonstrated by a number of groups [13-15]. Cote and his coworkers had investigated the ultracold atom-ion collision dynamics, charge transportation processes, and had realized possible formation of the stable atom-ion system [12]. It was shown recently that the elastic scattering cross section of an atom-ion system depends on the collisional energy in the semiclassical regime and favors scattering at small angles [16]. Particularly, the dispersion coefficients of the interacting inert gases with the alkaline earth ions and with the ground state of Li have extensive applications for understanding pressure broadening [17-20] and transportation of atoms in the laboratory experiments [21-23]. Necessity of accurate knowledge of dispersion interaction coefficients for the Li[+] ion interacting with the inert gas atoms at ultralow temperature is advocated in Refs. [12, 24]. Also, these van der Waals interactions are useful in estimating the refractive indices of the matter (atomic) waves traversing through the inert gases [25-28]. Values of the dispersion coefficients of these systems are essential for deducing the amount of pressure broadenings to estimate uncertainties accurately in the measured quantities [19, 20, 29, 30]. These coefficients can be used to manipulate the characteristics of potential surfaces of the amalgamated materials [31, 32].

When an ion is submerged in a buffer gas, the dispersion interaction gives rise to shifts in the transition frequencies between different atomic states. The above approach, which is used to develop the dispersion coefficients [17, 33-36], is pioneered by Dalgarno who has given these expressions in terms of oscillator strength sum rules [37-39]. Mitroy and co-workers had evaluated dispersion coefficients for Sr[+] ion by constructing one electron model of Sr[+] ion using semiempirical core potential [29]. Dispersion coefficients of Li, Li[+] and Be[+] interacting with rare gases have been described in detail in [40] by using electric dipole (E1) matrix elements, obtained employing a variational Hylleraas method, in a sum-over-states approach. Jiang *et al.* [41] had deduced dynamic polarizabilities and dispersion coefficients for the alkali atoms and for their ions using Casimir-Polder relations at the imaginary frequencies. In our previous work [42], we had determined the long range $c_6$ and $c_8$ coefficients among the alkali atoms and singly charged alkaline earth ions more accurately. Here, we extend these calculations further to a wide range of systems such as for the interacting inert gases with the alkali and alkaline ions, and for alkali atoms-alkali ion combinations. The analysis of the above dispersion coefficients requires evaluation of the dynamic electric dipole and quadrupole polarizabilities at the imaginary frequencies. In fact, the determination of accurate values of atomic polarizabilities itself has enormous applications in the areas of quantum information processing, optical cooling and trapping schemes and for studying atomic clocks [43]. For this purpose, we use all order relativistic coupled-cluster (RCC) and random phase approximation (RRPA) many-body methods to evaluate both the dipole and quadrupole polarizabilities of the considered systems. All the results are reported in atomic units (au) throughout the paper.

## 2 Theory of dispersion coefficients

Using the second order perturbation theory, the two body long-range dispersion interaction potential, with the interatomic separation distance $R$, can be expressed as [1, 33, 37, 38, 44, 45]

$$V_{disp}(R) = -\frac{c_6}{R^6} - \frac{c_8}{R^8}, \tag{1}$$

where the terms containing higher power denominator than $R^{-8}$ are neglected. The $c_n$ parameters (with $n$ =6, 8) are the van der Waals dispersion coefficients, where $c_6$ elucidates dipole-dipole interaction and $c_8$ represents for the dipole-quadrupole interactions between two atoms or between an atom and an ion combination [44, 46]. The coefficients $c_6$ and $c_8$ for two interacting systems $A$ and $B$ can be estimated using the expressions [47-49]

$$c_6 = \frac{3}{\pi} \int_0^\infty \alpha_1^A(i\omega) \alpha_1^B(i\omega) \, d\omega \tag{2}$$

and

$$c_8 = \frac{15}{2\pi} \left[ \int_0^\infty \alpha_1^A(i\omega) \alpha_2^B(i\omega) \, d\omega + \int_0^\infty \alpha_2^A(i\omega) \alpha_1^B(i\omega) \, d\omega \right], \tag{3}$$

where $\alpha_1^{A(B)}(i\omega)$ and $\alpha_2^{A(B)}(i\omega)$ are the respective dynamic dipole and quadrupole polarizabilities with imaginary frequency $(i\omega)$ for the system $A$ ($B$). These quantities for frequency $\omega$ can be written using the second order perturbation theory as given by

$$\alpha_k(\omega) = -\sum_{I \neq n} \frac{(E_n - E_I)|\langle \Psi_n|O_k|\Psi_I\rangle|^2}{(E_n - E_I)^2 - \omega^2}, \tag{4}$$

where $|\Psi_n\rangle$ represents for the ground state wave function, the sum over $|\Psi_I\rangle$ describes all the possible allowed excited states and $E$'s are the energies of their respective states. Here, $O_1$ is the electric dipole (E1) operator $D$ and $O_2$ is the electric quadrupole (E2) operator $Q$. In the next section, we discuss about the RCC and RRPA methods for determining the above dipole and quadrupole polarizabilities.

## 3 The RCC and RRPA methods of polarizabilities

We use the Dirac-Coulomb (DC) Hamiltonian in our calculations, which is given by

$$H_{DC} = \sum_i [c \vec{\alpha}_i \cdot \vec{p}_i + \beta_i c^2 + V_n(r_i)] + \sum_{\substack{i,j \\ i \geq j}} \frac{1}{r_{ij}}, \tag{5}$$

where $\vec{\alpha}$ and $\beta$ are the Dirac matrices and $V_n(r)$ is the nuclear potential. This is a good approximation to describe the positive energy states of the Dirac theory. Weak coupling with the positron wave functions are usually neglected and also the rest mass energy of the electrons can be subtracted for the convenience. Thus, the working DC Hamiltonian yields

$$H_{DC} = \sum_i [c \, \vec{\alpha_i} \cdot \vec{p}_i + (\beta_i - 1) c^2 + V_n(r_i)] + \sum_{\substack{i,j \\ i \geq j}} \frac{1}{r_{ij}}. \tag{6}$$

Again, it may not be appropriate to assume atomic nucleus as a point like object for accurate calculations. On the other hand, there are not proper valid models available to describe the nuclear structure exactly. Among many, Fermi charge distribution model is more popular in which density of an electron within the atomic nucleus is described by

$$\rho_n(r) = \frac{\rho_0}{1 + e^{(r-b)/a}}, \tag{7}$$

where $\rho_0$ is the normalization factor, $b$ is known as half-charge radius and $a = 2.3/(4 \ln 3)$ is related to the skin thickness of the nucleus. Considering this distribution, the nuclear potential can be obtained as

$$V_n(r) = \frac{Z}{\aleph \, r} \begin{cases} \frac{1}{b} \left( \frac{3}{2} + \frac{a^2 \pi^2}{2b^2} - \frac{r^2}{2b^2} + \frac{3a^2}{b^2} P_2 + \frac{6a^3}{b^2 r} (S_3 - P_3^+) \right) & \text{for} \quad a \leq b \\ \frac{1}{r} \left( 1 + \frac{a \, 62 \, \pi^2}{b^2} - \frac{3a^2 r}{b^3} P_2^- + \frac{6a^3}{63 b} (S_3 - P_3^-) \right) & \text{for} \quad r > b, \end{cases} \tag{8}$$

for the factors $\aleph = 1 + \frac{a^2 \pi^2}{b^2} + \frac{6a^3}{b^3} S_3$ with $S_k = \sum_{m=1}^{\infty} \frac{(-1)^{m-1}}{m^k} e^{-b/a}$ and $P_k^{\pm} = \sum_{m=1}^{\infty} \frac{(-1)^{m-1}}{m^k} e^{\pm m(r-b)/a}$. The $b$ parameter can be determined from $b = \sqrt{\frac{5}{3} r_{rms}^2 - \frac{7}{3} a^2 \pi^2}$ with the root mean square radius $r_{rms}$, which can be estimated using the empirical formula $r_{rms} = 0.836 A^{1/3} + 0.57$ in fermi (fm) or can be taken from a standard nuclear data table.

Owing to the two-body nature of the Coulomb interactions, solving eigenvalue equation for the atomic Hamiltonian $H_{at}$ given by

$$H_{at} |\Psi_n^{(0)}\rangle = E_n |\Psi_n^{(0)}\rangle, \tag{9}$$

with more than three electrons in an atomic system is infeasible. Instead, it is a usual practice to get the approximated solution to the above equation and then append corrections from the residual contributions gradually. This approximated solution is treated as a model space in the working Hilbert or Fock space accounting majority of the contributions from the Coulomb interactions in the calculation of the atomic wave functions. One of the most conducive and appropriate approaches to determine the approximated wave functions is to use the Hartree-Fock (Dirac-Fock (DF) in the relativistic framework) Hamiltonian ($H_0$). The residual interaction ($V_{res} = H_{at} - H_0$) can further improve the results by annexing contributions from the rest of the Hilbert or Fock space, referred to as orthogonal space, through a decent many-body method. Below we demonstrate few methods and try to inculcate one-to-one connections among these methods. For this purpose, we try to build-up each many-body approach by commencing from same DF wave function. To proceed further, we adopt the procedure of the generalized Bloch equation to explain the many-body methods systematically in a comprehensible and logical manner. In the many-body perturbation theory (MBPT) the exact wave function of an atomic state can be expressed as

$$|\Psi_n^{(0)}\rangle = \Omega_n^{(0)} |\Phi_n\rangle, \tag{10}$$

where $|\Phi_n\rangle$ is the model space (here DF wave function) and $\Omega_n^{(0)}$ is the wave operator which is responsible for incorporating contributions from the orthogonal space due to $V_{res}$. It should be noted that we also make here no-pair approximation while constructing the orthogonal space, i.e. the excited state configurations are built up considering only the zeroth order positive energy states, to avoid spurious contributions due to the contamination of the excited state configurations with the negative energy states. Contributions from the above orthogonal space can either be expressed in terms of order of perturbation or in the form of excited configurations with respect to $|\Phi_n\rangle$. Without loss of generality, we can go on with the perturbation series expansion approach first and then we can manifest the same in terms of the excited state configurations.

Two projection operators $P$ and $Q$ satisfying $|\Phi_n\rangle = P|\Psi_n^{(0)}\rangle$ and $Q = I - P$ for the identity operator $I$ are defined for easy description, which follows $P = |\Phi_n\rangle\langle\Phi_n|$. In the perturbative approach, it yields

$$\Omega_n^{(0)} = \Omega_n^{(0,0)} + \Omega_n^{(1,0)} + \Omega_n^{(2,0)} + \cdots = \sum_k \Omega_n^{(k,0)}. \tag{11}$$

Notice that we use two superscripts, for the later use, among which the first one represents for number of $V_{res}$ present in the calculations while the second one with zero means there is no external source of perturbation taken into account. The amplitudes of the above wave operators are solved one-by-one in the sequence of order of perturbations involved with the wave operators using the following recursive relation

$$[\Omega_n^{(k,0)}, H_0]P = QV_{res}\Omega_n^{(k-1,0)}P - \sum_{m=1}^{k-1} \Omega_n^{(k-m,0)}PV_{res}\Omega_n^{(m-1,0)}P. \tag{12}$$

The energy of the state ($E_n$) can be evaluated using an effective Hamiltonian $H_n^{eff} = PH\Omega_n^{(0)}P$ at different orders of perturbation with the expansion form of $\Omega_n^{(0)}$. i.e. $E_n = \langle\Phi_n|H_n^{eff}|\Phi_n\rangle$.

Now, the modified wave function ($|\Psi_n\rangle$) of the atomic system in the presence of an external weak perturbative source (such as $V_{prt}$ which can be either the $D$ or $Q$ operator for the evaluation of the dipole $\alpha_n^{E1}$ or quadrupole $\alpha_n^{E2}$ polarizabilities, respectively) can be approximated to first order approximation as

$$|\Psi_n\rangle = |\Psi_n^{(0)}\rangle + \lambda|\Psi_n^{(1)}\rangle, \tag{14}$$

where $\lambda$ is an arbitrary parameter representing the strength of the perturbation source. In this way, $\alpha_n^{E1}$ and $\alpha_n^{E2}$ can be obtained by expressing

$$\alpha_n^{E1/E2} = \frac{\langle\Psi_n|D|\Psi_n\rangle}{\langle\Psi_n|\Psi_n\rangle} \simeq \frac{\langle\Psi_n^{(0)}|D|\Psi_n^{(1)}\rangle}{\langle\Psi_n^{(0)}|\Psi_n^{(0)}\rangle}, \tag{15}$$

by considering $V_{prt} \equiv D$ or $V_{prt} \equiv Q$.

It is commanding to obtain solution for $|\Psi_n^{(1)}\rangle$ by solving an inhomogeneous equation of the type

$$(H_n^{eff} - E_n)|\Psi_n^{(1)}\rangle = (E_n^1 - V_{prt})|\Psi_n^{(0)}\rangle. \tag{16}$$

In Bloch equation methodology, we can express $|\Psi_n^{(1)}\rangle = \Omega_n^{(1)}|\Phi_n\rangle$ such as $\Omega_n^{(1)} = \sum_k \Omega_n^{(k,1)}$ encompassing $k^{th}$ order of $V_{res}$ and one order external perturbation $V_{prt}$. The amplitudes of $\Omega_n^{(1)}$ are obtained from the following equation

$$[\Omega_n^{(k,1)}, H_0]P = Q[V_{prt}\Omega_n^{(k,0)} + V_{res}\Omega_n^{(k-1,1)}]P -$$

$$\sum_{m=1}^{k-1}(\Omega_n^{k-m,0}PV_{prt}\Omega_n^{(m,0)}P - \Omega_n^{(k-m,1)}PV_{res}\Omega_n^{(m,0)})P. \tag{17}$$

For the choice of reference state $|\Phi_n\rangle$ as the DF wave function and external perturbation operator $V_{prt}$ being an one-body operator, the zeroth order expressions for the wave operators can yield $\Omega_n^{(0,0)} = 1$, $\Omega_n^{(1,0)} = 0$ and $\Omega_n^{(0,1)} = \sum_{p,a} \frac{\langle p|V_{prt}|a\rangle}{\epsilon_p - \epsilon_a}$ for the occupied $a$ and unoccupied $p$ orbitals with energies $\epsilon_a$ and $\epsilon_p$ respectively. In the double perturbative sources, up to $k = 0,1,2\cdots$ approximations in the wave operators are referred to MBPT(k) method.

Having said and done with the basic formalism of determining atomic wave functions in the many-body perturbative analysis, extending them to build-up these wave functions containing all orders in $V_{res}$ for both the cases, absence and presence of external source, would be now much straightforward. This can be achieved by generalizing the above perturbative approaches after carefully formulating the wave operator $\Omega_n$ in a slight different form or assembling the coefficients from each order of perturbation expansion to compose various degree of excitations. We now proceed to describe the RCC and RPA methods for calculating the polarizabilities. The all order perturbative nature of these many-body methods in determining the atomic wave functions can be understood well in the following manner.

In the RCC method, linear combination of the Slater determinants are carried out in a distinct manner so that atomic wave functions are contrived to form an exponential function. By assembling different level of excitations with respect to the DF wave function from each order of correction from the perturbation theory, we express the atomic wave function as

$$|\Psi_n^{(0)}\rangle = |\Phi_n\rangle + \sum_I^{N_I} C_n^I |\Phi_n^I\rangle + \sum_{II}^{N_{II}} C_n^{II} |\Phi_n^{II}\rangle + \sum_{III}^{N_{III}} C_n^{III} |\Phi_n^{III}\rangle + \cdots$$

$$\equiv |\Phi_n\rangle + \sum_I^{N_I} T_I^{(0)} |\Phi_n\rangle + \sum_{II}^{N_{II}} T_{II}^{(0)} |\Phi_n\rangle + \sum_{III}^{N_{III}} T_{III}^{(0)} |\Phi_n\rangle + \cdots$$

$$= |\Phi_n\rangle + T_1^{(0)} |\Phi_n\rangle + T_2^{(0)} |\Phi_n\rangle + T_3^{(0)} |\Phi_n\rangle + \cdots$$

$$= e^{T_1^{(0)} + T_2^{(0)} + T_3^{(0)} + \cdots + T_N^{(0)}} |\Phi_n\rangle = e^{T^{(0)}} |\Phi_n\rangle, \tag{18}$$

where $T^{(0)} = \sum_K^{N_K} T_K^{(0)}$ for $k = 1,2,3\cdots$ represents for RCC excitation operators with subscript $k$ implying $k^{th}$ level excitation carried out from $|\Phi_n\rangle$. The advantage of this method is of many fold: (i) it is both conceptually and computationally simpler, (ii) truncated RCC methods also satisfy both size-extensivity and size-consistency properties, (iii) owing to exponential form of the expression for the wave function, contributions from higher level excitations to certain extent also do appear through the non-linear terms in a truncated RCC method, etc.

Although we mentioned above about computational simplicity in the use of RCC method, yet in actual practice it does not turn out to be factual. Because of presence of the non-linear terms and requirement of a sufficiently large Hilbert or Fock space to carry out accurate calculations of the atomic wave functions, intermediate computational strategies have been adopted conforming available computational resources and depending upon the size of the atomic system of our interest. This is judiciously accomplished by devising a proper plan before implementing the method. Since atomic orbitals are meticulously described in the spherical coordinate system, thus use of reduced matrix elements instead of actual matrix elements were pertinent and it prevails extra computations for the azimuthal quantum numbers. This is the most well versed approach for states having closed-sell configurations, but states of open-shell configurations cannot be dealt with this way. However, atomic states having one or two electrons in the valence orbitals and one or two electron less from closed-shell configurations can be computed using the reduced matrix elements by appending valence orbitals or removing electrons from the appropriate closed-shell configurations in the Fock-space approach. We discuss here an approach to calculate wave functions of atomic sates having only one electron in the valence orbital along with a closed-shell configuration in the Fock-space formalism.

In the Fock-space CC formalism, wave functions of one valence ($v = n$) atomic states are expressed as

$$|\Psi_v\rangle = e^{T^{(0)}} \left\{ 1 + S_v^{(0)} \right\} |\Phi_v\rangle, \tag{19}$$

where $S_v$ is a CC operator exciting the valence electron $v$ along with from closed-core of $|\Phi_v\rangle$. In a Fock-space approach, $|\Phi_v\rangle$ is constructed from the closed-core $|\Phi_0\rangle$ by appending the respective valence orbital $v$ as $|\Phi_v\rangle = a_v^+ |\Phi_0\rangle$. In this approach, the RCC $T^{(0)}$ operator is responsible for accounting electron excitations from the closed-core $|\Phi_0\rangle$. In these expressions, superscript (0) is used to highlight that wave functions are still free from the external fields. We consider only the singles and doubles excitations in our calculations, which is known as the CCSD approximation in the literature.

The matrix element of an operator $O$ (which is either $D$ or $Q$ in or case) between the $|\Psi_f\rangle$ and $|\Psi_i\rangle$ states is determined by

$$\langle O \rangle_{fi} = \frac{\langle \Psi_f | O | \Psi_i \rangle}{\sqrt{\langle \Psi_f | \Psi_f \rangle \langle \Psi_i | \Psi_i \rangle}} = \frac{\langle \Phi_f | \{1+S_f^{(0)+}\} \bar{O} \{1+S_i^{(0)}\} | \Phi_i \rangle}{\sqrt{\langle \Phi_f | \{1+S_f^{(0)+}\} \bar{N} \{1+S_f^{(0)}\} | \Phi_f \rangle \langle \Phi_i | \{1+S_i^{(0)+}\} \bar{N} \{1+S_i^{(0)}\} | \Phi_i \rangle}}, \quad (20)$$

where $\bar{O} = e^{T^{(0)+}} O e^{T^{(0)}}$ and $\bar{N} = e^{T^{(0)+}} e^{T^{(0)}}$ are two non-truncated series in the above expression. These non-truncated series are computed at several intermediate steps in an iterative procedure.

We evaluate many E1 and E2 matrix elements among the ground and low-lying excited states using the above RCC method in the considered alkali atoms and singly ionized alkaline earth ions. Using these matrix elements and experimental energies, we estimate the dominant contributions to the dipole and quadrupole polarizabilities in a sum-over-states approach as given in Eq. (4). Other contributions such as from the core-valence correlations and high-lying excited states are estimated using the DF method as their magnitudes are negligibly small. The core correlations to these polarizabilities for the open-shell systems and the dipole and the quadrupole polarizabilities of the ground states of the closed-shell atomic systems are determined using the RRPA method, where the pair correlation effects contribute insignificantly. We outline the RRPA method briefly below.

The RRPA is a subclass of RCC method, but technically it is derived from the DF method in a completely different procedure. Its main advantage is, it can embody the core polarization effects to all orders at the same time being cost-effective. Its expression can be obtained from Eq. (12) by continuing $k$ to infinite order for $\Omega^{(k,1)}$ while suppressing $\Omega^{(k,0)}$ in a self-consistent procedure. The derivation of the final expression is a repercussion of expanding the DF wave function $|\Phi_n\rangle$ to first order due to $V_{prt}$ and generalizing it to infinite order. Thence, it only picks up the singly excited configurations from $|\Phi_n\rangle$ in case of polarizability calculations owing to one-body form of the interaction operator $V_{prt} \equiv D$ or $V_{prt} \equiv Q$. In the RRPA approach, the first order corrected wave operator $\Omega_n^{(1)} \equiv \Omega_n^{RPA}$ is explicitly given by

$$\Omega_n^{RPA} = \sum_{k=1}^{\infty} \sum_{pq,ab} \left\{ \frac{[\langle pb|V_{res}|aq\rangle - \langle pb|V_{res}|qa\rangle]\Omega_{b\to q}^{(k-1,1)}}{\epsilon_p - \epsilon_a} + \frac{\Omega_{b\to q}^{(k-1,1)+}[\langle pq|V_{res}|ab\rangle - \langle pq|V_{res}|ba\rangle]}{\epsilon_p - \epsilon_a} \right\}, \quad (21)$$

where $a \to p$ implies singly excitation operation by the wave operator replacing an orbital $a$ by $p$ in $|\Phi_n\rangle$. After obtaining amplitude of the $\Omega_n^{RPA}$ operator and using Eq. (15), we obtain the dipole and quadrupole polarizability contributions to the core correlations in the single valence atomic systems and ground state of the inert gas configured state functions.

## 4 Results

Our estimated dynamic polarizabilities for the alkali atoms and singly charged alkaline earth ions, obtained using the above discussed RCC method, are given in [42] and their accuracies are justified by comparing the static values with their corresponding experimental results. To verify how accurately we have also achieved these values in the inert gas atoms, we present our DF and RRPA results for these atoms in Table 1 and compare them with the results available from the other calculations and from the measurements. As can be seen from the table the differences between the DF and RRPA results are not substantial, however the RRPA calculations are closer with the experimental results. Johnson *et al.* [50] had also evaluated $\alpha_1$ values of the above inert gas atoms using RRPA, but considering different basis functions. Our results are very much consistent with their results. This implies that our RRPA can also provide accurate dynamic dipole and quadrupole polarizabilities for these atoms.

Soldan and colleagues had also employed a coupled-cluster (CC) method, in the non-relativistic framework, to determine these values [51] and had obtained more accurate results compared with the experimental values. It is very difficult to determine the dynamic polarizabilities for a large set of imaginary frequencies using the (R)CC methods due complexity involved in calculating the first order wave function due to the dipole and quadrupole operators. Nevertheless, our RRPA results are suitable enough for estimating the dispersion coefficients within the present interest of accuracy. There are no experimental results for $\alpha_2$ available to compare against our calculations, but we find few other theoretical studies on them using a variety of many-body methods [18, 41, 54]. Chen and co-workers [18] had evaluated the $\alpha_2$ values for the ground state of helium using a variational perturbative approach considering the B-spline and Slater-type basis functions in the configuration-interaction (CI) scheme. Thakkar *et.al* [54] had used a finite order many-body perturbation theory to gauge these quantities and in Ref. [41], Jiang and companions computed values of $\alpha_2$ using systematically generated effective oscillator strength distributions. Our results for $\alpha_2$ are consistent with these calculated results.

Table 1 **Calculated values of the static dipole and quadrupole polarizabilities for the He, Ne, Ar, Kr and Xe inert gases.**

| Polarizabilities | He | Ne | Ar | Kr | Xe |
|---|---|---|---|---|---|
| $\alpha_1^{total}$ (DF) | 0.99 | 1.98 | 10.15 | 15.82 | 26.87 |
| $\alpha_1^{total}$ (RRPA) | 1.32 | 2.37 | 10.77 | 16.47 | 26.97 |
| $\alpha_1^{total}$ (Other) | 1.32 [50], 1.383 [51] | 2.38 [50], 2.697 [52] | 10.77 [50], 11.22 [52] | 16.47 [50], 16.8 [52] | 26.97 [50], 27.06 [52] |
| $\alpha_1^{total}$ (Experiment) | 1.3838 [53] | 2.668 [53] | 11.091 [53] | 16.74 [53] | 27.340 [53] |
| $\alpha_2^{total}$ (DF) | 1.80 | 4.76 | 37.19 | 69.91 | 151.88 |
| $\alpha_2^{total}$ (RRPA) | 2.33 | 6.42 | 50.12 | 94.25 | 205.12 |
| $\alpha_2^{total}$ (Other) | 2.445 [18] | 7.52 [54] | 51.86 [54] | 98.2 [41] | 213.7 [41] |

Table 2 **Calculated values of the static dipole and quadrupole polarizabilities for the $Li^+$, $Na^+$, $K^+$ and $Rb^+$ alkali metal ions.**

| Polarizabilities | $Li^+$ | $Na^+$ | $K^+$ | $Rb^+$ |
|---|---|---|---|---|
| $\alpha_1^{total}$ (DF) | 0.16 | 0.83 | 5.46 | 9.27 |
| $\alpha_1^{total}$ (RRPA) | 0.19 | 0.95 | 5.45 | 9.06 |
| $\alpha_1^{total}$ (Other) | 0.1894 [50], 0.192486 [55] | 0.9457 [50], 1.00 [56] | 5.457 [50], 5.52 [56] | 9.076 [50], 9.11 [56] |
| $\alpha_1^{total}$ (Experiment) | 0.188 [57] | 0.978 [58] | 5.47 [58] | 9.0 [59] |
| $\alpha_2^{total}$ (DF) | 0.09 | 1.21 | 12.92 | 28.11 |
| $\alpha_2^{total}$ (RRPA) | 0.11 | 1.52 | 16.25 | 35.35 |

We have also performed the DF and RRPA calculations of polarizabilities for the singly charged alkali atoms. We present the static dipole and quadrupole polarizability values of these ions in Table 2 and compare them with the other available results. We find that the differences between the DF and the RRPA results are not very large and see that the RRPA results are much closer to their respective experimental values than in the case of the previously discussed inert gas atoms. This may be because of the fact that atomic orbitals are tightly bound in the ions than the neutral atoms. Our RRPA $\alpha_1$ values match well with the other RRPA values of Johnson *et al.* [50]. Lim and co-workers had also employed a RCC method considering only the scalar relativistic Hamiltonian [56] to evaluate these quantities, but their values are larger than the RRPA and experimental results. The reason could be their approximated method may be overestimating the correlation effects from the non-RRPA contributions and the higher order relativistic corrections may bring back these results closer to the experimental values. In this view and for the computational simplicity, it seems RRPA is appropriate to estimate the dynamic $\alpha_1$ values in these ions within the reasonable accuracies. We could not find any other references presenting the $\alpha_2$ results of these ions, but on the basis of findings on $\alpha_1$ values, we also assume that our RRPA calculations for $\alpha_2$ are of moderate accurate and can be considered for the accurate estimate of the dispersion coefficients.

Having gauged about the accuracies of our calculated polarizabilities, we now intend to determine the $c_6$ and $c_8$ dispersion coefficients for all possible combinations of the considered inert gases with the alkali and alkaline earth ions and the alkali atoms with their singly charged ions. Using the formula given by Eqs. (2) and (3) and the Gaussian quadrature integration method, we present the $c_6$ and $c_8$ values of the alkali ions with the inert gas atoms in Table 3. Tang and co-workers [40] had also evaluated these $c_6$ and $c_8$ coefficients for $Li^+$ interacting with the inert gases by using a variational Hylleraas method. It can be seen that the results obtained in the present work are comparative to the results calculated by the other groups.

Table 3 The $c_6$ and $c_8$ dispersion coefficients for the alkali ions-inert gases combinations. Values reported in Ref. [40] are also given to compare with our results.

| System | This work | | Others [40] | |
|---|---|---|---|---|
| | $c_6$ | $c_8$ | $c_6$ | $c_8$ |
| $Li^+$- He | 0.34 | 2.20 | 0.302715247023 | 1.9954705321 |
| $Li^+$- Ne | 0.72 | 5.86 | 0.66588 | 5.7756 |
| $Li^+$- Ar | 2.24 | 31.03 | 1.8427 | 27.256 |
| $Li^+$- Kr | 3.11 | 51.38 | 2.5468 | 45.261 |
| $Li^+$- Xe | 4.55 | 95.23 | 3.6392 | 83.937 |
| $Na^+$- He | 1.61 | 14.12 | | |
| $Na^+$- Ne | 3.23 | 35.06 | | |
| $Na^+$- Ar | 10.48 | 171.11 | | |
| $Na^+$- Kr | 14.64 | 261.11 | | |
| $Na^+$- Xe | 21.47 | 503.70 | | |
| $K^+$- He | 6.97 | 88.84 | | |
| $K^+$- Ne | 13.77 | 203.11 | | |
| $K^+$- Ar | 47.55 | 956.11 | | |
| $K^+$- Kr | 67.25 | 1534.64 | | |
| $K^+$- Xe | 100.08 | 2746.82 | | |
| $Rb^+$- He | 10.69 | 161.82 | | |
| $Rb^+$- Ne | 20.95 | 359.63 | | |
| $Rb^+$- Ar | 73.61 | 1655.53 | | |
| $Rb^+$- Kr | 104.52 | 2633.81 | | |
| $Rb^+$- Xe | 156.20 | 4661.33 | | |

Table 4 The $c_6$ and $c_8$ dispersion coefficients for the alkaline earth ions-inert gases combinations. Values reported in Ref. [41] are also given to compare with our results.

| System | This work | | Others [41] | |
|---|---|---|---|---|
| | $c_6$ | $c_8$ | $c_6$ | $c_8$ |
| $Ca^+$- He | 20.02 | 670.07 | 19.724 | 658.87 |
| $Ca^+$- Ne | 37.27 | 1319.37 | 39.136 | 1395.8 |
| $Ca^+$- Ar | 152.22 | 6243.18 | 145.07 | 5965.9 |
| $Ca^+$- Kr | 225.74 | 9879.45 | 212.77 | 9353.4 |
| $Ca^+$- Xe | 355.93 | 17263.05 | 329.07 | 16181 |
| $Sr^+$- He | 24.96 | 896.84 | 24.404 | 913.41 |
| $Sr^+$- Ne | 46.59 | 1753.64 | 48.613 | 1920 |
| $Sr^+$- Ar | 188.80 | 8283.49 | 178.02 | 8117.7 |
| $Sr^+$- Kr | 279.27 | 13072.11 | 260.40 | 12655 |
| $Sr^+$- Xe | 439.03 | 22723.46 | 401.31 | 21682 |
| $Ba^+$- He | 32.99 | 1428.28 | 31.504 | 1396.9 |
| $Ba^+$- Ne | 61.80 | 2771.32 | 62.753 | 2905.2 |
| $Ba^+$- Ar | 247.97 | 12790.04 | 229.44 | 12141 |
| $Ba^+$- Kr | 366.03 | 20004.12 | 335.37 | 18809 |
| $Ba^+$- Xe | 574.01 | 34315.91 | 516.40 | 31891 |
| $Ra^+$- He | 34.39 | 1462.77 | | |
| $Ra^+$- Ne | 64.80 | 2846.98 | | |
| $Ra^+$- Ar | 255.86 | 13060.71 | | |
| $Ra^+$- Kr | 375.96 | 20386.60 | | |
| $Ra^+$- Xe | 586.23 | 30873.79 | | |

Table 5 The $c_6$ and $c_8$ dispersion coefficients for the alkali atom-alkali ion combinations.

| System | $c_6$ | $c_8$ |
|---|---|---|
| Li- Li$^+$ | 3.28 | 157.21 |
| Li- Na$^+$ | 16.21 | 815.17 |
| Li- K$^+$ | 90.18 | 4773.52 |
| Li- Rb$^+$ | 148.14 | 8131.05 |
| Na- Li$^+$ | 3.87 | 189.23 |
| Na- Na$^+$ | 19.01 | 980.05 |
| Na- K$^+$ | 103.48 | 5728.29 |
| Na- Rb$^+$ | 169.21 | 9740.79 |
| K- Li$^+$ | 6.15 | 353.35 |
| K- Na$^+$ | 30.04 | 1810.07 |
| K- K$^+$ | 160.74 | 10527.01 |
| K- Rb$^+$ | 261.63 | 17782.86 |
| Rb- Li$^+$ | 7.12 | 445.85 |
| Rb- Na$^+$ | 34.71 | 2285.70 |
| Rb- K$^+$ | 183.63 | 13220.86 |
| Rb- Rb$^+$ | 297.98 | 22313.61 |

Similarly, we have tabulated the values of $c_6$ and $c_8$ coefficients for all the considered alkaline earth ion-inert gase combinations in Table 4. We find the magnitudes of these dispersion coefficients for the alkaline earth metal ions with the inert gases are less than the values of the corresponding alkali metal atom-inert gas interactions as the polarizabilities of the alkaline earth metal ions are smaller in magnitude. In Table 5, we present the $c_6$ and $c_8$ coefficients of the alkali metal atom-alkali metal ion combinations. Lee and co-workers [60], in year 2013, had estimated the collision rate coefficients for the collisions between Rb atoms and optically dark Rb$^+$ ions in trapped mixtures. Charge transfer and total cross sections in the elastic collisions of Na-Na$^+$ at ultralow temperatures have been studied by Cote and co-workers [11]. These dispersion coefficients may be useful in such experiments carried out using hybrid atom-ion traps [60].

## 5 Conclusion

In the foregoing work, we have determined the dynamic electric dipole and quadrupole polarizabilities in the alkali atoms and singly charged alkaline earth-metal ions using matrix elements that were obtained by employing the relativistic coupled-cluster. Similarly, a relativistic random phase approximation was used to calculate these quantities in the inert gas atoms and in the singly ionized alkali atoms and doubly ionized alkaline earth-metal atoms. Accuracies of these quantities were verified by comparing their static values with the other available theoretical and experimental results. By using these values, we determined the dispersion coefficients for the considered atomic systems, which are the constituent of long range interactions between the atoms or atom-ion combinations. These values will have significant applications in finding position of the magnetic field induced Feshbach resonances and to study collisional physics; particularly at the low-energy and low-temperature regime.

## 6 Acknowledgements

S.S. acknowledges financial support from UGC-BSR scheme. The work of B.A. is supported by CSIR grant no. 03(1268)/13/EMR-II, India. K.K. acknowledges the financial support from DST (letter no. DST/INSPIRE Fellowship/2013/758). B.K.S acknowledges use of Vikram-100 HPC Cluster at Physical Research Laboratory, Ahmedabad.

**References**
[1] M. Marinescu, H. R. Sadeghpour, and A. Dalgarno, Phys. Rev A **49**, 982 (1994).
[2] J. L. Roberts, N. R. Claussen, J. P. B. Jr, C. H. Greene, E. A. Cornell, and C. E. Wieman, Phys. Rev. Lett. **81**, 5109 (1998).
[3] D. M. Harber, J. M. McGuirk, J. M. Obrecht, and E. A. Cornell, J. Low Temp. Phys. **133**, 229 (2003).
[4] Y. Lin, I. Teper, C. Chin, and V. Vuletic, Phys. Rev. Lett. **92**, 050404 (2004).

[5] A. E. Leanhardt, Y. Shin, A. P. Chikkatur, D. Kielpinski, W. Ketterle, and D. E. Pritchard, Phys. Rev. Lett. **90**, 100404 (2003).
[6] P. J. Leo, C. J. Williams, and P. S. Julienne, Phys. Rev. Lett. **85**, 2721 (2000).
[7] M. Saffman, T. G. Walker, and K. Molmer, Rev. Mod. Phys. 82 **82**, 2313 (2010).
[8] C. Joachim, J. K. Gimzewski, and A. Aviram, Nature **408**, 541 (2000).
[9] K. Ravi, S. Lee, A. Sharma, G. Werth, and S. Rangwala, Nature Communications **3**, 1126 (2012).
[10] F. H. J. Hall and S. Willitsch, Phys. Rev. Lett. **109**, 233202 (2012).
[11] R. Cote and A. Dalgarno, Phys. Rev. A **62**, 012709 (2000).
[12] R. Cote, V. Kharchenko, and M. D. Lukin, Phys. Rev. Lett. **89**, 093001 (2002).
[13] A. Rakshit and B. Deb, Phys. Rev. A **83**, 022703 (2011).
[14] L. Ratschbacher, C. Zipkes, C. Sias and M. Kohl, Nat. Phys. **8**, 649 (2012).
[15] S. Willitsch, Proc. Int. Sch. Phys. Enrico Fermi **189**, 255 (2015).
[16] A. Harter and J. H. Denschlag, Contemporary Physics **55**, 33 (2014).
[17] C. Zhu, A. Dalgarno, S. G. Porsev, and A. Derevianko, Phys. Rev. A **70**, 032722 (2004).
[18] M. K. Chen and K. T. Chung, Phys. Rev. A **53**, 1439 (1996).
[19] G. Peach, Adv. Phys. **30**, 367 (1981).
[20] N. Allard and J. Kielkopf, Rev. Mod. Phys. **54**, 1103 (1982).
[21] L. A. Viehland and D. S. Hampt, J. Chem. Phys. **97**, 4964 (1992).
[22] M. F. McGuirk, L. A. Viehland, E. P. F. Lee, W. H. Breckenridge, C. D. Withers, A. M.Gardner, R. J. Plowright, and T. G.Wright, J. Chem. Phys. **130**, 194305 (2009).
[23] A. M. Gardner, C. D. Withers, T. G. Wright, K. I. Kaplan, C. Y. N. Chapman, L. A. Viehland, E. P. F. Lee, and W. H. Breckenridge, J. Chem. Phys. **132**, 054302 (2010).
[24] U. E. Senff and P. G. Burton, Mol. Phys. **58**, 637 (1986).
[25] C. Champenois, E. Eudouard, P. Duplaa, and J. Vigue, J. Phys. II **7**, 523 (1997).
[26] S. Blanchard, D. Civello, and R. C. Forrey, Phys. Rev. A **67**, 013604 (2003).
[27] T. D. Roberts, A. D. Cronin, D. A. Kokorowski, and D. E. Pritchard, Phys. Rev. Lett. **89**, 200406 (2002).
[28] J. Mitroy and J.-Y. Zhang, Phys. Rev. A **76**, 032706 (2007).
[29] J. Mitroy, J. Y. Zhang, and M. W. Bromley, Phys. Rev. A **77**, 032512 (2008).
[30] H. Harima, K. Tachibana and Y. Urano, Phys. Rev. A **35**, 109 (1987).
[31] S. S. Xantheas, G. S. Fanourgakis, S. C. Farantos, , and M. Velegrakis, J. Chem. Phys. **108**, 46 (1998).
[32] D. Prekas, B.-H. Feng, and M. Velegrakis, J. Chem. Phys. **108**, 2712 (1998).
[33] J. -Y. Zhang and J. Mitroy, Phys. Rev. A **76**, 022705 (2007).
[34] J. Mitroy and J. Y. Zhang, Eur. Phys. J. D **46**, 415 (2008).
[35] A. Derevianko, W. R. Johnson, M. Safronova, and J. Babb, Phys. Rev. Lett. **82**, 3589 (1999).
[36] S. G. Porsev and A. Derevianko, J. Chem. Phys. **119**, 844 (2003).
[37] A. Dalgarno and W. D. Davison, Adv. At. Mol. Phys. **2**,1 (1966).
[38] A. Dalgarno, Adv. Chem. Phys. **12**, 143 (1967).
[39] T. M. Miller and B. Bederson, Adv. At. Mol. Phys. **13**,1 (1977).
[40] L. Y. Tang, J. Y. Zhang, Z. C. Yan, T. Y. Shi, and J. Mitroy, J. Chem. Phys. **133**, 104306 (2010).
[41] J. Jiang, J. Mitroy, Y. Cheng, and M. W. J. Bromley, At. Data and Nucl. Data Tables **101**, 158 (2015).
[42] J. Kaur, D. K. Nandy, B. Arora, and B. K. Sahoo, Phys. Rev. A **91**, 012705 (2015).
[43] W. M. Itano, J. Res. NIST **105**, 829 (2000).
[44] J. M. Standard and P. R. Certain, J. Chem. Phys. **83**, 3002 (1985).
[45] M. Marinescu, J. F. Babb, and A. Dalgarno, Phys. Rev A **50**, 3096 (1994).
[46] B. K. Sahoo, Chem. Phys. Lett. **448**, 144 (2007).
[47] P. Kharchenko, J. F. Babb, and A. Dalgarno, Phys. Rev. A **55**, 3566 (1997).
[48] J. Mitroy, M. S. Safronova, and C. W. Clark, J. Phys. B **43**, 202001 (2010).
[49] B. Arora and B. K. Sahoo, Phys. Rev. A **89**, 022511 (2014).
[50] W. R. Johnson, D. Kolb, and K. Huang, At. Data Nucl. Data Tables **28**, 333 (1983).
[51] P. Soldan, E. P. F. Lee, and T. G. Wright, Phys. Chem. Chem. Phys. **3**, 4661 (2001).
[52] T. Nakajima and K. Hirao, Chem. Lett. **30**, 706 (2001).
[53] P. W. Langhoff and M. Karplus, J. Opt. Soc. Am. **59**, 863 (1969).
[54] A. J. Thakkar, H. Hettema, and P. E. S. Wormer, J. Chem. Phys. **97**, 3252 (1992).
[55] A. K. Bhatia and R. J. Drachman, Can. J. phys. **75**, 11 (1997).
[56] I. S. Lim, J. K. Laerdahl, and P. Schwerdtfeger, J. Chem. Phys. **116**, 172 (2002).
[57] W. E. Cooke, T. F. Gallagher, R. M. Hill, and S. A. Edelstein, Phys. Rev. A **16**, 1141 (1977).
[58]H. Eissa, U. Opik, Phys. Soc. **92**, 566 (1967).
[59] I. Johansson, Ark. Fys. **20**, 135 (1961).
[60] S. Lee, K. Ravi, and S. A. Rangwala, Phys. Rev. A **87**, 052701 (2013).